\documentstyle[12pt,aasms4]{article} 
\def\etal{{et~al.}\ }
\def\vol#1  {{{#1}{\rm,}\ }}

\def\lya{{\rm Ly}\alpha}
\def\mytau{\tau_{{\rm Ly}\alpha}}

\def\etal{et al.\ }

\def\clock{\count0=\time \divide\count0 by 60
     \count1=\count0 \multiply\count1 by -60 \advance\count1 by \time
     \number\count0:\ifnum\count1<10{0\number\count1}\else\number\count1\fi}

\begin{document}
\title{Testing Cosmological Models With A $\lya$ Forest Statistic: 
The High End Of The Optical Depth Distribution}
\author{Renyue Cen}
\vskip 0.5cm
\centerline{Princeton University Observatory, Princeton, NJ 08544}
\vskip 0.5cm
\centerline{cen@astro.princeton.edu}
%\vskip 0.7cm
%\centerline{draft of Aug 15, 1996}

\begin{abstract}

We pay particular attention to the high end of the $\lya$
optical depth distribution of a quasar spectrum. 
Based on the flux distribution (Miralda-Escud\'e \etal 1996),
a simple yet seemingly 
cosmological model-differentiating statistic, $\Delta_{\tau_0}$
--- the cumulative probability of a quasar spectrum 
with $\lya$ optical depth greater than
a high value $\tau_0$ --- is emphasized.
It is shown that two different models ---
the cold dark matter 
model with a cosmological constant and the mixed hot and cold
dark matter model, both normalized to COBE 
and local galaxy cluster abundance ---
yield quite different values of $\Delta_{\tau_0}$: 
0.13 of the former versus 
0.058 of the latter for $\tau_0=3.0$ at $z=3$.
Moreover,
it is argued that $\Delta_{\tau_0}$
may be fairly robust to compute theoretically 
because it does not seem to depend sensitively on 
small variations of simulations parameters such as 
radiation field, cooling, feedback process, radiative
transfer, resolution and simulation volume
within the plausible ranges of the concerned quantities.
Furthermore, it is illustrated that 
$\Delta_{\tau_0}$
can be obtained sufficiently accurately from currently
available observed quasar spectra for $\tau_0\sim 3.0-4.0$,
when observational noise is properly taken into account.
We anticipate that
analyses of observations of quasar $\lya$ absorption spectra 
over a range of redshift may be able to 
constrain the redshift evolution of the amplitude 
of the density fluctuations on small-to-intermediate scales,
therefore providing an independent constraint on
$\Omega_0$, $\Omega_{0,HDM}$ and $\Lambda_0$.

\end{abstract}


\section{Introduction}

Any acceptable theory for growth of structure
has to pass the tests imposed by observations
in our local ($z=0$) universe.
Among those the most stringent is provided by
observations of clusters of galaxies
(\cite{bc92};
\cite{ob92};
\cite{bc93};
\cite{wef93};
\cite{vl95};
\cite{bm96};
\cite{ecf96};
Pen 1996),
simply because they probe the
tail of a Gaussian (or alike) distribution,
which depends extremely strongly on some otherwise
fairly stable quantities such as the 
rms of density fluctuation amplitude.
In addition, a model has to match 
COBE observations of the universe
at the epoch of recombination (\cite{s92}).
The combination of
these two relatively fixed points defines the shape
and amplitude of the assumed power spectrum,
thus significantly limiting the allowable parameter space
for the Gaussian family of the cold dark matter cosmological models;
one is essentially left undecided
how to adjust the following three parameters:
$n$, $\Lambda_0$ or $\Omega_{0,hot}$, 
where $n$ is the primordial power spectrum index,
$\Lambda_0$ the current cosmological constant and
$\Omega_{0,hot}$ density parameter of the hot dark matter.
Critical differentiators are likely to come from 
areas which are as far removed as possible from both COBE epoch 
(on large scales)
and our local vantage point (on scales of $\sim 8h^{-1}$Mpc).

In the intervening redshifts,
among those accessible to current observations,
the $\lya$ forest observed in spectra of high
redshift quasars 
(e.g., Carswell \etal 1991; Rauch \etal 1993; Petitjean \etal 1993;
Schneider \etal 1993; Cristiani \etal 1995; Hu \etal 1995;
Tytler \etal 1995;
Lanzetta \etal 1995; Bahcall \etal 1996)
provides possibly the single richest set of information about
the structure of the universe at low-to-moderate redshift.
Furthermore, each line of sight to a distant quasar 
indiscriminately samples the
distribution of neutral hydrogen gas (hence total gas)
over a wide redshift range ($z\sim 0-5$) in a 
{\it random} fashion 
(i.e., a quasar and foreground absorbing material are unrelated);
thus, the $\lya$ forest constitutes perhaps 
the {\it fairest} sample of the
cosmic structure in its observed redshift range.

The new {\it ab initio} modeling of 
$\lya$ clouds by following the 
gravitational growth of baryonic density fluctuations 
on small-to-intermediate scales 
($\sim 100 $kpc to several Mpc in comoving length units)
produced a remarkably successful account 
of the many observed properties of 
$\lya$ clouds
(Cen \etal 1994; Zhang \etal 1995; Hernquist \etal 1996;
Miralda-Escud\'e \etal 1996). 
However, at first sight it appears that
three {\it different} cosmological models ---
a CDM+$\Lambda$ model (Hubble constant 
$H_o=65$km~s$^{-1}$~Mpc$^{-1}$, 
$\Omega_{0,CDM}=0.3645$, $\Lambda_0=0.6$,
$\Omega_{0,b}=0.0355$, $\sigma_8=0.79$; Cen \etal 1994),
a biased critical density CDM model (
$H_o=50$km~s$^{-1}$~Mpc$^{-1}$, 
$\Omega_{0,CDM}=0.95$, $\Omega_{0,b}=0.05$, 
$\sigma_8=0.70$; Hernquist \etal 1996),
and a high amplitude critical 
density CDM model ($H_o=70$km~s$^{-1}$~Mpc$^{-1}$, 
$\Omega_{0,CDM}=0.96$, $\Omega_{0,b}=0.04$,
$\sigma_8=1.00$; Zhang \etal 1995) 
--- all approximately fit observations, 
gauged by column density
distribution, Doppler width $b$ distribution, etc.
While the overall agreement between models
and observations is encouraging from an astrophysicist's point of view
because it implies that the physical
environment produced by the hydrocodes of widely distributed photoionized
gas does correspond to the real world,
it seems that more sensitive tests are demanded from a
cosmologist's point of view in order to test models.
In this {\it Letter} we show that 
a statistic based on the $\lya$ flux distribution
may serve as a potentially
strong discriminator between cosmological models.
We argue that, while the column density distribution is useful in
providing information about the $\lya$ clouds,
it requires post-observation fitting procedures such as line
profile fitting and line deblending, and 
it consequently superimposed by 
additional uncertainties related to the
procedures themselves.
Naively, it may seem that
measures based on the flux (or optical depth) distribution
may be redundant given that we already have the 
traditional column density distribution.
But we note that in complex multivariate
problems (Kendall 1980)
such as that of the $\lya$ forest
it should not be taken for granted that a one-to-one
correspondence or correlation between the two exists.
In other words, two different flux distributions
may yield a similar column density distribution
considering the many factors involved,
including real space clustering, thermal broadening, peculiar velocity 
effect and line profile fitting.
Therefore, it may be profitable
to deal directly with the flux distribution.

\section{A Statistic: Fraction Of A Spectrum With $\tau_\lya > \tau_0$}

We use two of the current popular models --- the cold dark matter 
model with a cosmological constant ($\Lambda$CDM) and the mixed hot and cold
dark matter model (MDM) --- to demonstrate 
the applicability of the statistic.

We begin by showing the variance of the density fluctuations 
($\sigma\equiv\sqrt{\langle\delta^2\rangle-1}$,
where 
$\delta\equiv\delta\rho/\langle\rho\rangle$ calculated using linear theory)
as a function of the comoving radius of a sphere
in the $\Lambda$CDM model (solid curve)
and the MDM model (dashed curve) at $z=3$ in Figure 1.
Both the $\Lambda$CDM model (Hubble constant 
$H_o=65$km~s$^{-1}$~Mpc$^{-1}$, 
$\Omega_{0,CDM}=0.3645$, $\Lambda_0=0.6$,
$\Omega_{0,b}=0.0355$, $\sigma_8=0.79$)
and the MDM model 
($H_o=50$km~s$^{-1}$~Mpc$^{-1}$, 
$\Omega_{0,CDM}=0.74$, $\Omega_{0,HDM}=0.2$,
$\Omega_{0,b}=0.06$, $\sigma_8=0.70$; Ma 1996)
are normalized to both COBE on large scales 
and
local galaxy cluster abundance (on scales of $\sim 8h^{-1}$Mpc comoving).
Both models involve a slight tilt of the spectrum (plus
some small gravitational wave contribution to the temperature fluctuations
on large scales in the $\Lambda$CDM model) in order to achieve
the indicated $\sigma_8$ values.
We see a somewhat {\it modest} difference in the amplitude 
of density fluctuations in the two models.
The density fluctuations are larger in the $\Lambda$CDM than
in the MDM model by (26\%, 33\%) on scales of 
($0.1h^{-1}$Mpc, $1.0h^{-1}$Mpc), respectively.

Focusing on the high end
of the optical depth distribution 
the following statistic is examined for 
the purpose of testing cosmological models ---
the fraction of pixels in a quasar spectrum
with $\lya$ optical depth greater than $\tau_0$:
\begin{equation}
\Delta_{\tau_0} \equiv P(>\tau_0),
\end{equation}
\noindent 
where $P(>\tau_0)$ is the cumulative distribution of $\lya$ optical depth.
In Figure 2, we show the results of the $\Lambda$CDM model 
(thick solid curve) and the MDM model (thick dashed curve)
obtained from detailed synthetic $\lya$ spectra
(Miralda-Escud\'e \etal 1996). 
A sampling bin of 2.0km~s$^{-1}$ is used and 
the spectrum is degraded by a point spread function
with a 6.0km~s$^{-1}$ FWHM.
Note that noise is {\it not} added for the two thick curves in Figure 2.
The boxsize, resolution and physics input of 
the two simulations are identical (for details
of the $\Lambda$CDM model see Miralda-Escud\'e \etal 1996).
The differences between
the two model simulations are:
1) different input power spectra,
2) different background models and,
3) there is one additional (hot) dark matter species in the MDM simulation.
Ionization equilibrium between photoionization
and recombination is assumed.
We require that
the average flux decrement in each model, 
$\langle D\rangle=1-\langle \exp^{-\mytau}\rangle$,
be equal to the observed value, $\langle D\rangle_{obs}=0.36$ 
at $z=3$ (\cite{prs93}).
This is achieved by adjusting the following parameter: 
$\mu\equiv {(\Omega_{0,b}/0.015h^{-2})\over (h~j_{H}/10^{-12} \hbox{sec}^{-1})^{1/2}}$,
where $h$ is Hubble constant in units of $100$km/s/Mpc,
$\Omega_{0,b}$ is the mean baryonic density and
$j_{H}$ is the hydrogen photonization rate.
This normalization procedure for the
flux distribution is {\it necessary} in order to fix its overall amplitude,
due to the large uncertainties of the observed values of
$\Omega_{0,b}$ and $j_H$.
The fitted values of $\mu$ are $1.90$ and $1.47$,
respectively, for the $\Lambda$CDM and the MDM models
(note that the value of $\mu$ for the MDM model is obtained
after the temperature of the
intergalactic medium in the model is raised; see below).

Although
the $\Lambda$CDM model yields temperature of the clouds consistent 
with observations, as indicated by the $b$ parameter distribution
(Miralda-Escud\'e \etal 1996),
it is found that the mean temperature of the intergalactic medium
in the MDM model at $z=3$ is 
unrealistically low with $\langle T\rangle\sim 100$~Kelvin
(compared to $\langle T\rangle\sim 15,000$~Kelvin in the $\Lambda$CDM).
The reason is that our self-consistent treatment of structure
formation (star/galaxy formation) and ionizing radiation field
does not produce sufficient
photo-ionization/photo-heating of the intergalactic medium
due to the very small fraction of baryons which 
have collapsed to form stars or quasars in the MDM model by $z=3$
[for details of our treatment of atomic processes,
radiation, and galaxy formation see Cen (1992) 
and Cen \& Ostriker (1993)].
It would be meaningless had we generated 
the $\lya$ spectrum for the MDM model using such low temperature.
Instead, we uniformly raise the gas temperature in the MDM model
to $2\times 10^{4}$~Kelvin (but retain other properties
of the gas including density and velocity),
and then generate the $\lya$ spectrum with thermal broadening effect
of the gas of the putative high temperature.
To test the sensitivity of the 
results on the artificial temperature adjustment
in the MDM model, we also
compute the results by raising the temperature to 
$4\times 10^{4}$~Kelvin and $1\times 10^{4}$~Kelvin, respectively.
We find that at $\tau_0=(3.0,4.0,5.0)$,
the results are $\Delta_{\tau_0}=(0.054,0.039,0.029)$,
$(0.058,0.041,0.032)$ and 
$(0.071,0.051,0.041)$ for three
cases with $T=4\times 10^4$, 
$T=2\times 10^4$, 
$T=1\times 10^4$~Kelvin, respectively.
One more experiment is made for the MDM model:
instead of setting the temperature uniformly to 
$2\times 10^{4}$~Kelvin,  
$2\times 10^{4}$~Kelvin is {\it added} uniformly to
the temperature of each cell, 
and we find the results of
$\Delta_{\tau_0}$ for the two cases are indistinguishable
within the quoted digits, at all three $\tau_0$'s.
It seems that the results depend only weakly 
on the temperature within the reasonable range,
with the trend that higher temperatures yield lower
fractions of pixels with high optical depth.

We find that 
$\Delta_{\tau_0} (\hbox{$\Lambda$CDM})=(0.12,0.10,0.088)$ and
$\Delta_{\tau_0} (\hbox{MDM})=(0.058,0.040,0.031)$,
at $\tau_0=(3.0,4.0,5.0)$, respectively,
i.e., a difference between the two models by a factor 
of $(2.1,2.5,2.8)$ at the three $\tau_0$'s.
It is likely that 
the high end of the $\lya$ optical depth distribution
in the MDM model 
(thick dashed curve) would be further reduced, had we run
the simulation with sufficient (i.e. realistic)
photo-ionization/photo-heating,
since reduced cooling and increased thermal pressure
would result in less condensation of dense regions responsible
for high $\lya$ optical depth considered here.
The countervailing effect is that the current MDM 
simulation may have overcooled the dense clouds, making them
much smaller thus much
less capable of intercepting QSO lines of sight.
However, examinations of the cloud sizes and densities
in the dense regions as well as the found trend of 
higher temperatures yielding lower high optical depth fraction 
(see the preceding paragraph)
indicate that this effect is unimportant for the MDM model;
there do not seem to exist superdense clouds in the MDM model.
In short, we anticipate that a more realistic
MDM simulation with higher photoionization field
would produce even smaller 
$\tau_0$ than that of the current MDM simulation.

It is clear from Figure 2 that the higher the $\tau_0$,
the more model-differentiating
power $\Delta_{\tau_0}$ possesses.
However, detecting high $\tau_0$
is difficult due to 
noise and telescope systematics.
To see how noise affects the statistic,
we generated synthetic spectra with noise added in the following way.
By definition, the signal to noise ratio at the continuum is 
S/N$={N_{src}\over \sqrt{N_{src}+N_{noise}}}$,
where $N_{src}$, $N_{noise}$ and S/N are 
the number of source photons at the continuum,
the number of noise photons
and the signal to noise ratio, respectively, per frequency pixel.
Thus, given S/N and $N_{noise}$, we can obtain $N_{src}$.
To simplify the illustration 
(without loss of generality) we assume that 
$N_{noise}$ is dominated by the detector readout noise.
This is a good approximation only for a bright quasar
where the number of sky photons are small
(due to a shorter exposure time) compared to
the readout noise of, say, a CCD detector.
For example, the gain of the HIRES CCD detector
on the Keck telescope is 6.1 electrons, so the number of photons due to the
CCD readout noise integrated over 5 spatial pixels (for each frequency bin)
is $N_{CCD}=5\times 6.1^2=186$.
For a $V=16.5$ mag quasar at $5000$A with 1-hr integration time,
about 4 photons from the sky per spatial pixel,
giving a total count of sky photons per frequency pixel (integrated
over the 5 spatial pixels) of only 20 photons
(see, e.g., Hu \etal 1995).
A frequency pixel in the simulated noise-free spectrum with flux $f$
contains $f N_{src}$ photons.
When noise is added,
the ``observed" number of photons (subtracted by the
known CCD noise) in the pixel will be 
$Poisson(f N_{src} + N_{CCD}) - N_{CCD}$,
where $Poisson(X)$ means a Poisson distributed random number
with the mean equal to $X$.
The resultant distributions [$P(\tau)$] are also shown
in Figure 2 for three continuum signal to noise ratios
of S/N$=(50,100,150)$ for each model (thin solid curves for 
the $\Lambda$CDM 
model and thin dashed curves for the MDM model).
The $6\sigma$ statistical errorbars are also shown for
the case with S/N$=100$
(other cases have comparable errorbars but are
not shown to maintain the readability of the plot),
assuming a quasar absorption spectrum coverage
of unit redshift range about z=3 with a sampling bin of $2$km~s$^{-1}$.
The $N_{CCD}$ value of the HIRES CCD detector is adopted in the calculation.
We see that three values of S/N$=(50,100,150)$ 
are able to preserve the differences
between the two models up to $\mytau\sim (2.5,3.0,3.5)$,
respectively.
Another complication, telescope systematics,
may cause further difficulties.
Nevertheless,
it seems that 
$\Delta_{\tau_0}$ for 
$\tau_0=3.0$ can be 
fairly accurately obtained with a high statistical accuracy
in good Keck spectra 
[e.g., Womble, Sargent, \& Lyons (1995) achieved
a typical signal to noise ratio of 150 per resolution element
using the HIRES spectrograph on the Keck telescope].

A preliminary comparison of the simulation results using $\Delta_{\tau_0}$
with observations (Rauch \etal 1997, Figure 1)
at $\tau_0=3.0$ at $z \sim 3$ appears that the result of 
the $\Lambda$CDM model matches the observed value
well ($\sim 0.10$ computed versus $\sim 0.11$ observed),
while the MDM model seems to predict a value ($0.04$)
lower than observed.

\section{Discussion and Conclusions}

This study illustrates
that high quality quasar $\lya$ absorption spectra 
are potentially useful to discriminate between cosmological models.
It is demonstrated that the $\Delta_{\tau_0}$ statistic --- 
the cumulative probability of a spectrum with $\lya$ optical depth
greater than a high value $\tau_0$ ($\sim 3.0-5.0$) ---
serves as an amplifier of the differences between models.
We show that a modest difference ($\sim 25-30\%$)
in the mean amplitudes 
translates into a large difference in the two $\Delta_{\tau_0}$'s
(by a factor larger than 2.0 for $\tau_0>3.0$)
between the $\Lambda$CDM model and the MDM model at $z=3$.
Moreover, the value of $\Delta_{\tau_0}$ is
at the level of 0.01 to 0.1, i.e., the relevant regions with 
$\mytau > \tau_0$
cover a sizable portion of a quasar spectrum.
Therefore, one is {\it not} dealing with small number statistics,
ensuring that $\Delta_{\tau_0}$ is 
a potentially accurately determinable statistic
statistically.
The statistic is also {\it simple} in that
it does not involve complicated procedures such as
line profile fitting, and hence can be directly applied
to the observed flux distribution.

In addition to the need of a high signal-to-noise ratio
(S/N$\ge 100$ for $\tau_0\ge 3.0$), 
it is essential that telescope systematics be understood 
and scattered noise photons be minimized, 
which may push one to focus on the brightest quasars at this time.
Furthermore, it is required that 
the spectroscopic 
resolution be sufficiently high so that $\lya$ optical depth can be
reliably extracted from the observed flux.
The latter requirement is equivalent to having 
a FWHM less the Doppler width, which seems readily 
satisfied with current observations.
Finally, we note that  
a stable normalization procedure for the 
overall flux distribution is necessary
due to large uncertainties in $\Omega_{0,b}$ and $j_H$.
We adopt $\langle D\rangle_{obs}$ as a normalization parameter
in this work.
The primary difficulty in fixing 
$\langle D\rangle_{obs}$
is to determine the continuum, which
is fairly obscured by heavy absorption at high redshift
(compare, e.g., Zuo \& Lu 1993 and \cite{prs93} to see
the situation); the situation is better at lower redshift.

It is equally essential to determine the robustness
of the prediction of a theoretical model for the proposed statistic. 
This may only be made definitive by 
performing many simulations with varying input parameters
including the ionization
radiation field, baryonic density, feedback processes,
simulation resolution and simulation volume (boxsize).
It is expected, however, that all these effects 
may not change results significantly.
Let us take an example to illustrate this conjecture.
A $\lya$ cloud with a column density 
of $N=1\times 10^{14}$cm$^{-2}$
and a Doppler width $b=25$km~s$^{-1}$ has a central
$\lya$ optical depth of $\sim 3.0$.
Since these clouds are
not cooling efficiently (cooling time is longer
than the Hubble time),
%(usually adiabatic cooling and photoionization heating are balanced),
cooling effects are likely to be small,
implying that changing radiation field or baryonic density
would have a small dynamic effect within plausible ranges.
For the same reason, these clouds are not effective in forming
stars, therefore feedback effect may be small 
(which, if any, may be due to nearby star forming regions, which
may be much rarer.)
These clouds are also found to be relatively large and well resolved
in our current simulations, but small
compared to the simulation boxsize, so
an increase of simulation resolution or boxsize would not 
affect the statistic substantially.
Lastly, we note that, since the relevant regions are optically
thin to Lyman limit photons, self-shielding effects
are likely to be unimportant.
So it appears that $\Delta_{\tau_0}$ is a relatively
{\it easy} statistic to determine theoretically.

We anticipate that the evolution of different models
may be quite different,
due to the dependence of the growth of density fluctuations
on $\Omega_0$, $\Omega_{0,HDM}$ and $\Lambda_0$
(Peebles 1980), possibly coupled 
with other nonlinear, redshift dependent thermal/dynamical effects.
The redshift evolution 
of $\Delta_{\tau_0}$, computable 
with both observations and simulations
of different models, may be potentially revealing.
It is conceivable that we can 
constrain the amplitude of density
fluctuations on small-to-intermediate scales as a function
of redshift using observations of $\lya$ clouds,
thus possibly constrain $\Omega_0$, $\Omega_{0,HDM}$ and $\Lambda_0$.

\acknowledgments
I am grateful to Limin Lu for discussions on 
the observational procedure and for many stimulating conversations,
to Jordi Miralda-Escud\'e for allowing me to use his
synthetic spectrum program and for his critical reading of 
the first version of the manuscript.
I thank David Weinberg for a careful reading of and 
many instructive comments on the manuscript.
Discussions with Jerry Ostriker are also acknowledged.
The work is supported in part by grants NAG5-2759 and ASC93-18185.

\newpage

\figcaption[Figure 1]{
shows the variance of density fluctuations 
%($\sigma\equiv\sqrt{\langle\delta^2\rangle -1}$ with 
%$\delta\equiv\delta\rho/\langle\rho\rangle$, 
%calculated using linear theory)
as a function of scale (comoving)
in the $\Lambda$CDM model (solid curve)
and the MDM model (dashed curve)
at redshift $z=3$.
\label{fig1}}

\figcaption[Figure 2]{
shows the (cumulative) distributions of the optical depth
in the $\Lambda$CDM model 
(solid curve) and the MDM model (dashed curve) at $z=3$.
The two thick curves do not include observational noise.
The three thin solid curves include observational noise
with signal to noise ratio at the continuum 
being $(50,100,150)$, respectively, for the $\Lambda$CDM model. 
The three thin dashed curves show the counterparts for the MDM model. 
The $6\sigma$ statistical errorbars are also shown for
the case with S/N$=100$
(other cases have comparable errorbars but are
not shown for the sake of readabilty of the figure),
assuming that a quasar absorption spectrum coverage
of unit redshift range about z=3 with a sampling bin of $2$km~s$^{-1}$.
The value of $N_{CCD}=181$ of the HIRES CCD detector 
is adopted in the calculation of noise added spectra
(noise due to sky photons is ignored).
All curves are normalized to yield the observed average
decrement at z=3 of $\langle D\rangle_{obs}=0.36$ 
(Press, Rybicki, \& Schneider 1993).
\label{fig2}}

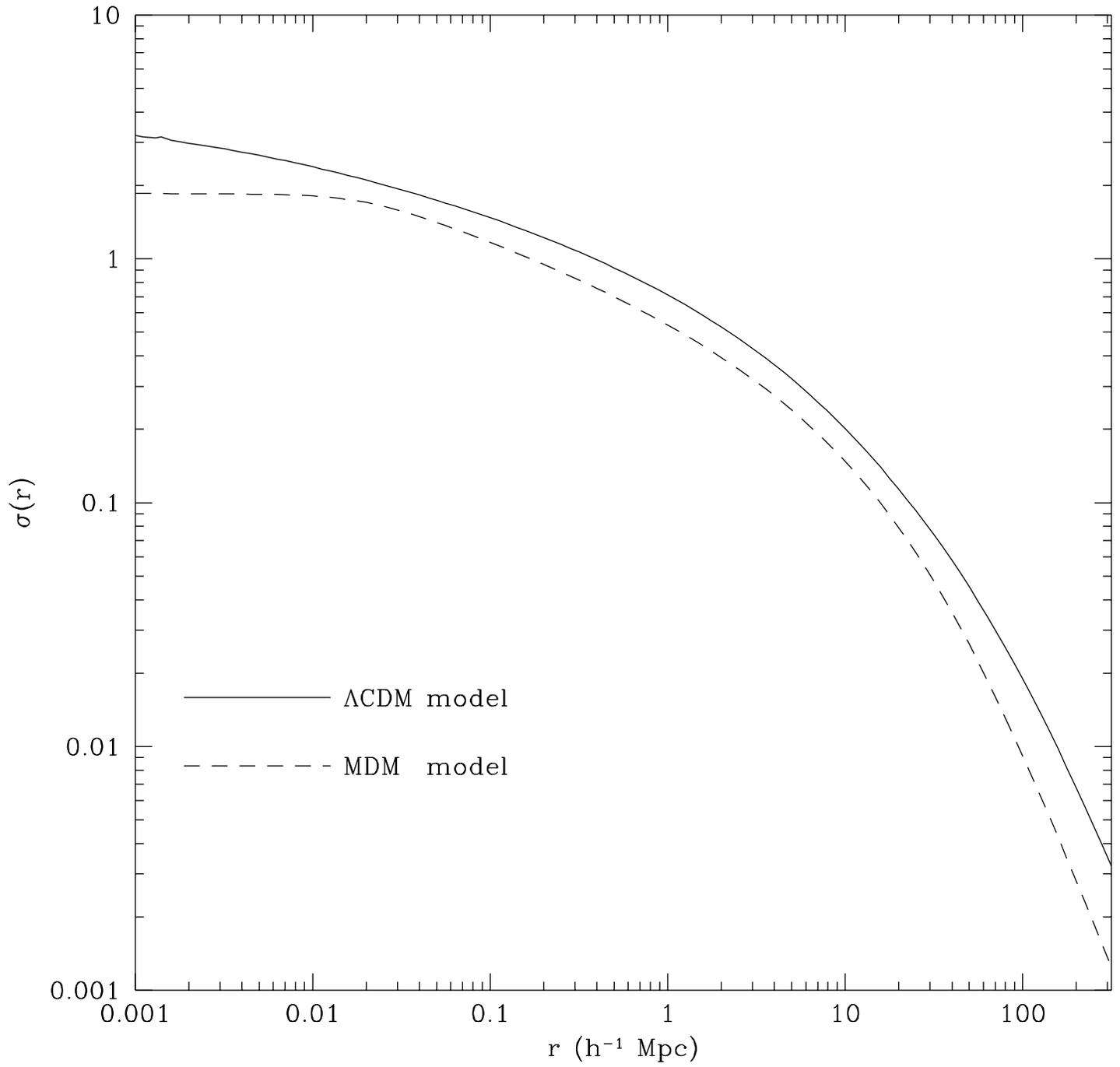
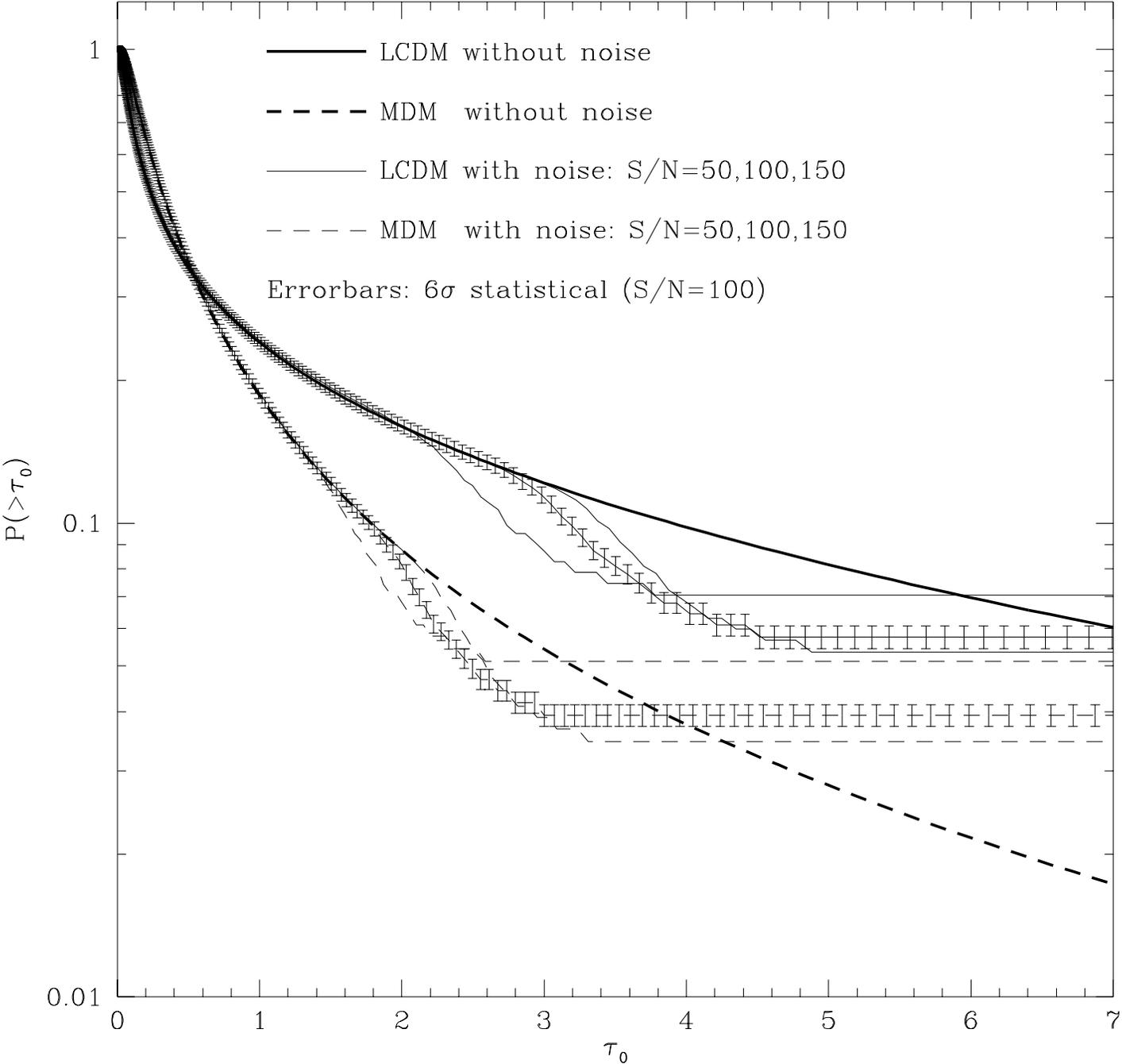
\end{document}